\documentclass[aip,amsmath,amssymb,reprint]{revtex4-1}
\usepackage{graphicx} 
\usepackage[T1]{fontenc}
\usepackage[utf8]{inputenc}

\begin{document}
\title{High-yield electrochemical etching of nanometrically defined Fe STM tips}
\author{Jędrzej Tepper}
\affiliation{Huygens-Kamerlingh Onnes Laboratory, Leiden University Niels Bohrweg 2, 2333CA Leiden}
\author{Jan M. van Ruitenbeek}
\affiliation{Huygens-Kamerlingh Onnes Laboratory, Leiden University Niels Bohrweg 2, 2333CA Leiden}

\date{\today}

\begin{abstract}
    A reproducible procedure for creating STM tips with nanometrically defined apices out of 0.25mm iron wire is presented.
\end{abstract}
\maketitle

\section{Introduction}
The quality of data obtained during scanning tunneling microscopy (STM) operation is restricted by the tip's properties.\cite{Binnig1982,Chen1992} This implies that even under ideal conditions and with a suitable sample, the experiment will not produce the expected results if the probe does not have the required characteristics for a specific measurement. Most STM measurements require a straight apex structure with a nanometrically defined radius. Electrochemical etching is the most commonly used way to reproducibly create STM probes with sufficiently small apices out of metal wires. A more recent development involves using magnetic tips to study local spin-dependent properties,\cite{Wiesendanger1990,Wulfhekel1999} Although there has been research on etching bulk 
iron tips,\cite{Haze2019} we found that the success rate of this procedure was insufficiently predictable. In this work, we focus on a specific high-yield procedure confirmed through detailed scanning electron microscopy (SEM) inspection. Throughout this study, we have identified several important factors that impact the success rate.

\begin{figure*}[ht!]
    \centering
    \includegraphics[width=1\linewidth]{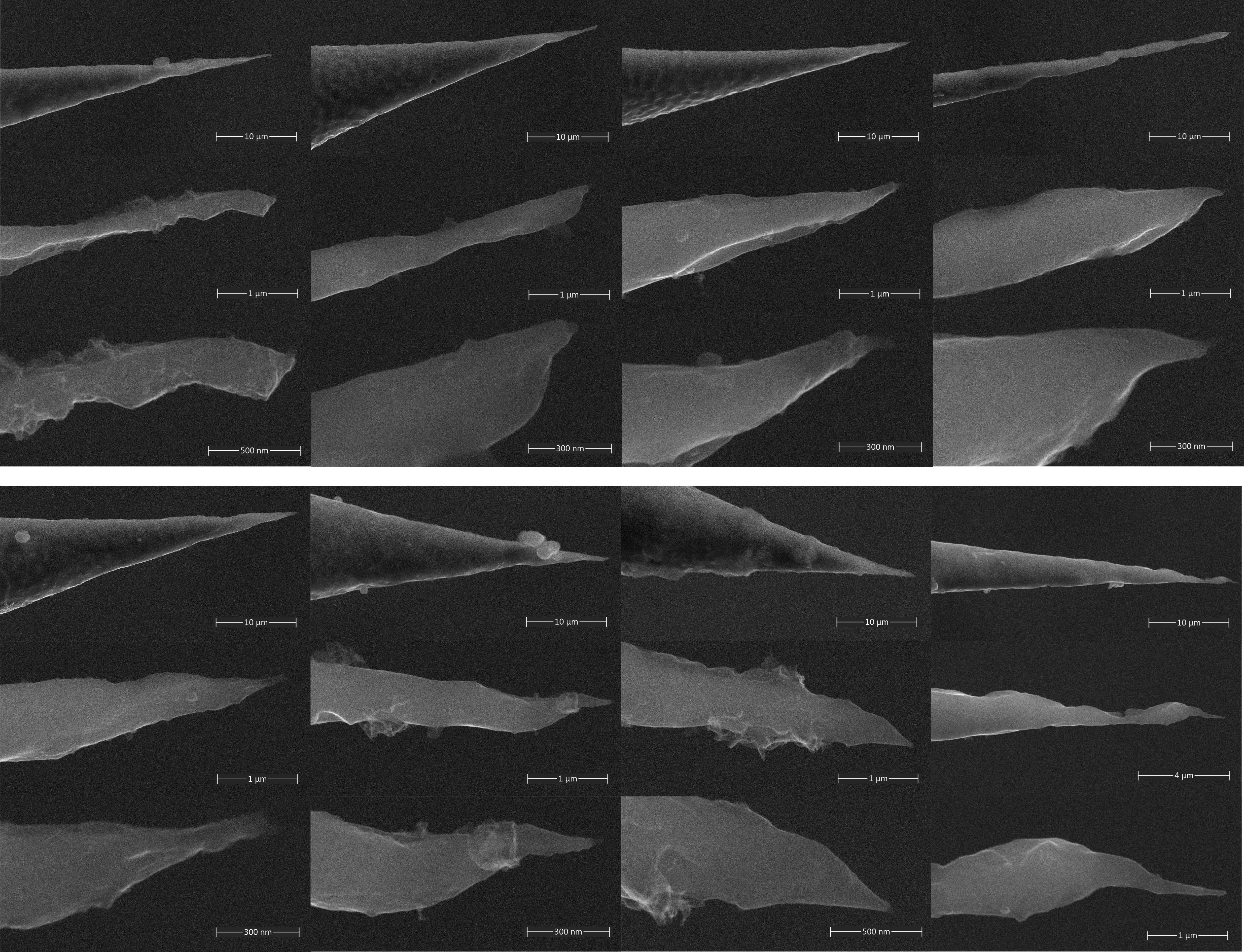}
    \caption{Scanning electron microscopy (SEM) images for a series of eight iron tips, etched one after another. Three different magnifications are presented for each specimen. }
    \label{fig:fe}
\end{figure*}

Our approach is based on the method widely used for tungsten tip preparation, the lamellae drop-off technique described by Klein \& Schwitzgebel.\cite{Klein1997} Following this work, we control the area of the electrochemical etching process by using a small volume of electrolyte suspended as a thin film inside a cathode shaped in the form of a ring. The weight of the wire suspended beneath the electrolyte film creates a gravity-driven stopping mechanism for the reaction, preventing blunting of the apex. Obtaining straight and nanometrically defined tip apices in a reproducible manner out of iron proved challenging and required slowing down the process in its final stage.

\section{The etching procedure}
Demineralized water should be used to prepare a 2M KCl solution. Ultrasonication for 15 minutes effectively ensures thorough dissolution, but alternatively, the solution can be heated. It is crucial to use an electrolyte free of particles such as dust to prevent the lamella from bursting during the etching process. This can be achieved by filtering the solution using Luer-lock syringes with disc PTFE 0.22{\textmu}m filters just before use.

A beaker with filtered 2M KCl is used to form a lamella suspended by its surface tension over a ring with an inner diameter of 9mm, made from 0.25mm >99\% gold wire, which forms the cathode of the electrochemical circuit. The anode is formed by a 0.25mm >99\% iron annealed wire, immersed into the lamella by piercing the wire through the center of the electrolyte. The length of the wire beneath the plane defined by the cathode ring is essential, as it determines the weight that ultimately tears the thinned wire apart. The length that ensures drop-off of the bottom part without premature tearing negatively influencing the apex radius was found to be 15mm $\pm$1mm.

Etching is initiated by applying a constant-current power supply with a voltage limit of 3V and a setpoint of 10mA. The voltage needed to sustain this current should start at $\leq$2V, indicating a sufficient volume of electrolyte and good electrical contacts between electrodes and the power supply. As the reaction proceeds, the voltage gradually rises, indicating an effective reduction of the area participating in the etching process. Eventually, the maximum voltage of 3V will not be high enough to sustain the initially set current. The decrease of the etching rate resulting from the reduced current in its last stages before the drop-off gives more time for any microscopical readjustments of the center of mass of the bottom part of the wire to take place. The typical etching process takes less than 5 minutes.

Candidates for good quality tips fall vertically onto a bed of shaving foam placed directly under the bottom of the anode. They are collected using tweezers and submerged three times in 80$^{\circ}$C demineralized water, a critically important step that removes most of the electrolyte remnants, followed by triple submersion in isopropanol to remove water.

\section{The drop-off}
The moment when the tip falls is determined by two concurrent processes: mechanical tearing of the wire caused by the weight of the lower end of the anode wire, and the rate of electrochemical etching. Ideally, the probe should drop in a straight line. To have an accessible experimental measure indicating how much lateral motion occurred during the drop-off, shaving foam is used as a collecting medium. When the fallen tips have their bottom part immersed into the foam, a significantly higher yield can be expected compared to when the tips are lying on it horizontally.

There are two main causes for non-vertical drop-offs: external forces acting on the wire beneath the lamella and forces stemming from the geometry and composition of the anode material. The latter are more difficult to avoid in an experiment and will therefore be described first.

As the etching of the wire proceeds and therefore its diameter is reduced, its rigidity is also reduced. The part of the anode beneath its thinned part can exhibit a damped pendulum-like movement just before the drop-off happens. If this motion is sufficient to tear the wire, the apex formed is often bent at a 
large angle, up to ~90$^{\circ}$, rendering the tip unusable. In an ideal situation, the thinnest part is perfectly aligned with the center of mass of the bottom part until the drop-off moment. In reality, this is never the case. The length of the bottom part constitutes leverage, creating a mechanical advantage for any lateral force originating from this part of the wire. The operator needs to align the bottom of the anode with the center of the electrolyte lamella. To facilitate this process, a mechanical point of support placed centrally above the cathode ring is used. This support should preferably be in the form of a tube inert to the electrochemical etching with a bore diameter that allows the anode wire to be guided inside without damage. 

External forces that disrupt the vertical drop-off can have various sources, depending on the environment where the etching apparatus is located. It was found that minor mechanical or acoustic vibrations do not affect the yield. However, drafts are an important source of lateral forces that lead to reduced yields. It is therefore recommended to fully enclose the etching setup to prevent the bottom part of the anode from moving due to air currents.

\section{The lamellae}

One of the most common problems encountered with the drop-off methods is the bursting of the lamella during the etching process. The origins of this problem are often elusive. We have found that particles such as dust locally change the surface tension of the lamella and therefore facilitate its bursting. This source of etching interruption can be avoided to a large degree by filtering the electrolyte right before use. Cathode material and geometry also play an important role, since these factors influence the volume of KCl solution that can be suspended. 0.25mm gold wire easily forms a symmetrical ring of 9mm inner diameter that is inert to the electrochemical process and holds an optimal electrolyte volume. Additionally, it was found that this material required only minor cleaning between consecutive probe etchings in the case of KCl electrolyte and iron anode. Whenever an etching process is finished, a cleaning tissue is used to remove the lamella along with the suspended products of the electrochemical reaction. Next, a beaker containing demineralized water is used to submerge the gold ring and dilute any uncollected remnants from the cathode. Once this cleaning is done, the lamella can be reestablished.

Eventually, however, the cathode will accumulate enough foreign material on its surface that bursting can happen despite using filtered electrolyte. A separate beaker with electrolyte can be used to fully immerse the cathode and pass the current of the order of 1A for a few seconds which will thoroughly clean its surface. After this process, a beaker with demineralized water should be used to dissolve most of the electrolyte remnants.

\section{Results}

Our procedure allowed us to successfully produce a set of 8 consecutively etched iron tips, shown in figure \ref{fig:fe}. All presented specimens fulfill the main prerequisite for STM probes, they have a straight and nanometrically defined apex, which has been confirmed through high-magnification SEM micrographs. This result proves that the main factors influencing the yield of the process are well controlled.

\section{Conclusions}
In conclusion, we have presented a highly reproducible method of etching STM tips out of iron wire. To achieve nanometrical apices in this metal, we needed to slow down the process in its final stage. This was achieved by driving the electrochemical reaction using a readily available constant current power supply with a voltage limit. The etched probes with the highest chance of being straight at the apex will fall vertically. To easily identify the path that the tip followed during falling, we used shaving foam as a collecting medium. Whenever the bottom of the tip is immersed vertically into the foam, yields are close to 100\%, making SEM inspection unnecessary and therefore the whole procedure quick and accessible.

\bibliography{references}

\end{document}